\documentclass[12pt]{article}\usepackage[dvips]{graphicx}
\usepackage{latexsym}
\usepackage{amssymb}

\newcommand{\C}{\mathbb{C}}
\newcommand{\CP}{\mathbb{CP}}

\newcommand{\R}{\mathbb{R}}

\newcommand{\Z}{\mathbb{Z}}

\renewcommand{\d}{\mathrm{d}}
\usepackage{amsmath, amsthm, amssymb}
\newtheorem{theorem}{Theorem}[section]

\def\be{\begin{equation}}
\def\ee{\end{equation}}

\def\p{\partial}
\def\ov{\overline}

\begin{document}
\title{\vskip -70pt
\begin{flushright}
{\normalsize DAMTP-2006-97} \\
\end{flushright}
\vskip 80pt
{\bf Moduli spaces with external fields}
\vskip 20pt}

\author{Maciej Dunajski\thanks{M.Dunajski@damtp.cam.ac.uk}\\[8pt]
{\sl Department of Applied Mathematics and Theoretical Physics} \\
{\sl University of Cambridge} \\
{\sl Wilberforce Road, Cambridge CB3 0WA, UK} \\[10pt]
and \\[10pt]
Marcin Ka\'zmierczak \thanks{marcin.kazmierczak@gmail.com} \\[8pt]
{\sl Department of Theoretical Physics}\\
{\sl University of Warsaw}\\
{\sl Ho\.za 69, 00--681 Warsaw, Poland}\\[10pt]
}
\date{} 
\maketitle
\begin{abstract}
{We consider the geometric structures on the moduli space of static
finite energy solutions to 
the $2+1$ dimensional unitary chiral model with the
Wess--Zummino--Witten (WZW) term. It is shown that the  magnetic
field induced by the WZW  term vanishes when restricted to the moduli spaces 
constructed from the Grassmanian embeddings, so that the slowly moving
solitons can in some cases be approximated by a geodesic motion on a space of
rational maps from $\CP^1$ to the Grassmanian.} 
\end{abstract}
\newpage
\section{Introduction}
\setcounter{equation}{0}
Let us consider a fairly general framework for field theory.
The {\it space time} $(M, \eta)$ is a $(D+1)$-dimensional 
manifold with a Lorentzian metric $\eta$, and the {\it target space} $(Y, h_Y)$
is  a $k$-dimensional
manifold with a (pseudo) Riemannian metric $h_Y$. The dynamics of the
theory depends on a choice of the action, which is a functional on
the space of maps $\mbox{Map}(M, Y)$.
In the  {\it canonical} approach one sets $M=\Sigma\times \R$, and
regards the field equations as the
infinite dimensional dynamical system on the space ${\cal M}$ of maps 
$J:\Sigma\longrightarrow Y$, where the initial data set $\Sigma$ is
the  $D$-dimensional 
manifold with a Riemannian metric induced by $\eta$. The details
depend on the model, but generally one aims to formulate the evolution
equations as the geodesic motion (possibly with a potential) on 
${\cal M}$. The $L^2$ metric on $\cal M$ is induced by the target space
metric $h_Y$ in the following way. For a given map $J$ we identify
$T_J{\cal M}$ with the space of maps $X:\Sigma\longrightarrow TY$ such
that $\pi\circ X=J$, where $\pi:TY\rightarrow Y$ is a natural projection, 
and set
\be
\label{metric_int}
|X|^2=\int_{\Sigma}h_Y(X(p), X(p)) {\rm{\>dp}},\qquad X\in T_J{\cal M},
\ee
where $ {\rm{\>dp}}$ is some measure on $\Sigma$.

In the case of gravity (where the overall structure differs slightly
from the one described above)
this leads to the exact procedure  realising 
the Einstein equation as a dynamical system, where the metric on 
${\cal M}$ is the celebrated deWitt metric, and the potential is
given by the scalar curvature of the Riemannian metric on the initial
data set $\Sigma$. To make it all work one needs to factor out $
{\cal M}$ by the action
of the group of diffeomorphisms of $\Sigma$, and ensure that the initial data
satisfies the constraint equations \cite{FM71}.

In gauge theory, where  one considers the quotient of ${\cal M}$ by
the infinite dimensional group of gauge transformations, the metric 
(\ref{metric_int}) coincides with the inner product induced by the
kinetic term in the Lagrangian on $M$. Here the
emphasis has been on the approximate techniques. In the moduli space
approximation \cite{M82} the dynamics is restricted to a finite
dimensional submanifold of ${\cal M}$. This submanifold consists of
appropriately chosen static
finite energy solutions to the full field equations. If the `initial
position' is given by a static solution which
minimises the potential energy of the field configuration and initial
kinetic energy is small, then the trajectory in  
${\cal M}$ can be expected to stay close to a geodesic in the manifold of 
static finite energy
solutions.
This `follows from' the total energy conservation. In some cases
the whole procedure can be made rigorous \cite{Stuart}. See \cite{MS}
for a review of the geodesic method.

In other cases (including various string theories in 10 dimensions and
11-dimensional super-gravity), the target space $Y$ admits a rich structure
consisting of more than just a (pseudo) Riemannian metric. In
particular any differential $(D+r)$-form on $Y$ induces an $r$ form on 
${\cal M}$ in a way which does not depend on $h_Y$. If this
differential form appears in the
Lagrangian it will give rise to an
external (magnetic like) field on the moduli space. This can lead
to interesting physical consequences. If the topology of
$\cal M$ is non-trivial the Aharonov-Bohm effect may take place 
on the space of solutions even if the magnetic field vanishes \cite{WZ86}.

In this paper we shall give a detailed analysis of one example where
an external field arise on the moduli space.
We shall take  the space-time $M$ to  be $\R^{2, 1}$, and the target space
to be a unitary Lie group with its natural trace form  metric. 
Any Lie
group  admits a connection which parallel propagates
left--invariant vector fields. This connection is flat, but necessarily
has torsion. Using this
connection with torsion in the chiral model Lagrangian 
modifies the equations of motion, and surprisingly makes them
integrable \cite{W88T}. This modification can also be  interpreted in
terms of the WZW term in the chiral model action. In the rest of this
section  we shall introduce 
this modified chiral model, originally due to Ward \cite{W88}. We shall
also review its static solutions given in terms of the Grassmanian
embeddings $\CP^n\longrightarrow U(n+1)$. 
In Section \ref{section2} we shall construct a metric and
a magnetic potential on the moduli space of static solutions. 
The corresponding magnetic field will be shown to vanish, but the 
flat magnetic connection can
still be interesting, since the moduli space (which in our case 
consists of based rational maps  $\CP^1\longrightarrow \CP^n$, where the
two-sphere $\Sigma=S^2=\CP^1$ is the initial data set compactified by the
boundary conditions)
is not simply connected. In Section \ref{section3} we shall show that
the magnetic 1-form can be obtained canonically from a pull back
of a certain 1-form from $\CP^n$. In the Appendix we discuss the Noether 
currents arising from the WZW Lagrangian.
Some of the results presented in this paper appeared in the MSc Thesis
of the second author.
\subsection{Modified chiral model}
Consider a smooth map $J:\R^{2,1}\longrightarrow U(n+1)$. The
integrable chiral model 
is defined by equation
\begin{equation}\label{wardeq}
\begin{aligned}
&\left({\eta^{\mu\nu}-V_\alpha\epsilon^{\alpha\mu\nu}}\right)\> 
(J^{-1}J_{\mu})_{\nu}=0 ,\\
&\eta={\rm{diag}}(-1,1,1) , \ \ \ \ V_\alpha=(0,1,0) , \ \ \ \ 
\epsilon^{012}=1 ,
\end{aligned}
\end{equation}
where Greek letters denote three dimensional space-time indices taking
values $0,1,2\equiv t,x,y$. 
The abbreviated notation of differentiation $J_{\mu}\equiv
\partial_{\mu}J$ and the summation convention is going to be used in 
the article. A choice of the unit space-like vector $V=\p/\p x$ breaks
the 
Lorentz invariance down to $SO(1, 1)$, but ensures the integrability
of (\ref{wardeq}).\par

The Lagrangian formulation of (\ref{wardeq}) contains the
Wess--Zumino--Witten (WZW) term \cite{torsion, IZ98}.  This involves an 
extended field 
$\hat{J}$ defined in the
interior of a cylinder which has the space--time as one of its
boundary components
\[
\hat{J}:\R^{2+1}\times[0,1]\longrightarrow U(n+1)
\]
such that $\hat{J}(x^{\mu}, 0)$ is a constant group element, which we
take to be the identity ${\bf 1}\in U(n+1)$, and
$\hat{J}(x^{\mu}, 1)=J(x^{\mu})$.

The equation (\ref{wardeq}) can be derived as a 
stationary condition for the
action functional
\begin{equation}\label{S}
\begin{aligned}
&S=S_C+S_M \ ,\\
&S_C=-\frac{1}{2}\int_{[t_1,t_2]\times
  \mathbb{R}^2}{\rm{Tr}}\left({\mathbb{J}\wedge 
\star \mathbb{J}}\right) \ , \\
&S_M=\frac{1}{3}\int_{[t_1,t_2]\times \mathbb{R}^2\times
  [0,1]}{\rm{Tr}}\left({\mathbb{\hat J}\wedge \mathbb{\hat J}\wedge
    \mathbb{\hat J}\wedge \mathbb{V}}
\right) \ ,
\end{aligned}
\end{equation}
where $J$ should be treated as a field.
Here $\star $ is a Hodge star of $\eta_{\mu\nu}$ and 
\[
\mathbb{J}=J^{-1}J_{\mu}\>dx^{\mu} \ , \ \ \ \ \mathbb{\hat J}=
\hat J^{-1} \hat J_{p}\>dx^{p} \ , \ \ \ \ p=0,1,2,3\equiv t,x,y,\rho
\]
are $\mathfrak{u}(n+1)$-valued 
 1-forms  on $\mathbb{R}^{2+1}$ and $\mathbb{R}^{2+1}\times [0,1]$
 respectively and $\mathbb{V}={\bf{1}}\,dx$ is a constant 1-form  
on $\mathbb{R}^{2+1}\times [0,1]$. 
We make an assumption that the extension $\hat J$ is of the form 
\begin{equation}\label{warunek}
\hat J(x^\mu,\rho)=\mathcal{F}\left({J(x^\mu),\rho}\right) 
\end{equation}
for some smooth function $\mathcal{F}:U(n+1)\times 
[0,1]\longrightarrow U(n+1)$.
The WZW term $S_M$ in the action is topological in the sense that its
integrand does not depend on the metric on $\R^{2,1}$.

Following \cite{witten} we can obtain a more geometric picture by 
regarding the domain of $\hat{J}$ as $B\times \R$, where $B$ is a ball
in  $\R^3$ with
the boundary $\p B=S^2$ regarded as a compactified space,  and 
rewriting
$S_M$ as 
\[
S_M=\int_{[t_1, t_2]\times B}\hat{J}^*(T)\wedge V \ , \ \ \ \ V=\d x \
 .
\]
Here $T$ is the preferred three--form  \cite{torsion} 
on $U(n+1)$ in the third cohomology group given by 
$T =\mbox{Tr}[(\phi^{-1} \d\phi)^3]$ for $\phi\in U(n+1)$. This three
form  coincides with 
torsion of a flat connection  $\nabla$ on $U(n+1)$ which parallel propagates
left--invariant vector fields, i.e.
\begin{equation*}
T(X,Y,Z)=g\left({\nabla_{X}Y-\nabla_{Y}X-[X, Y] \ , \ Z}\right)  ,
\end{equation*}
where $g=-\mbox{Tr}(\phi^{-1}\d \phi\; \phi^{-1}\d \phi)$
is the metric on $U(n+1)$ given in terms of the Maurer--Cartan one form 
(this definition
makes sense for any matrix Lie group).

The torsion three--form $T$ can  be pulled back to $B$. It is closed, so $T=\d
\lambda$, where $\lambda$ is a two--form on $G$ which can be defined 
only locally. The Stokes theorem now yields
\begin{eqnarray*}
S_M&=&\int_{[t_1, t_2]\times B}
\d (\hat{J}^*(\lambda)\wedge V)\\
&=&\frac{1}{2}\int_{S^2\times [t_1, t_2]}(\varepsilon^{\mu\nu\alpha}V_\alpha)
\lambda_{ij}(\phi)
\p_{\mu}\phi^i\p_{\nu}\phi^j\; \d x\;\d y\;\d t,
\end{eqnarray*}
where $\phi^i=\phi^i(x^{\mu})$ are local coordinates on the 
group (e.g the components of the matrix $J$). 
In the above derivation we have
neglected the boundary component $(t_1\times B)\cup (t_2\times B)$, as
variations of the  corresponding integrals vanish identically.

Time translational invariance of $S$ gives rise to the conservation of 
the energy functional which appears to be the same as for the ordinary
chiral
model\footnote{In the Appendix we shall construct the corresponding 
momenta in $x$ and $y$ directions.}
\begin{equation}\label{energy}
\begin{aligned}
E&=T+E_p \ ,\\
T&=-\frac{1}{2}\int_{\mathbb{R}^2}{\rm{Tr}}\left({(J^{-1}J_t)^2}\right)
{\rm{\>dx\>dy}}
\ , \\
E_p&=-\frac{1}{2}\int_{\mathbb{R}^2}{\rm{Tr}}\left({(J^{-1}J_x)^2
+(J^{-1}J_y)^2}\right){\rm{\>dx\>dy}}.  
\end{aligned}
\end{equation}
Finiteness of energy can be  ensured \cite{W88} by imposing the
boundary condition on $J$
\begin{equation}\label{fecond}
J(t,r,\theta)=J_0+r^{-1}J_1(\theta)+O(r^{-2}) \ , \qquad
x+iy=re^{i\theta} \ ,
\end{equation}
where $J_0$ is a constant matrix \cite{W88} and the whole dependence
on $t$ lies in $O(r^{-2})$.
\subsection{Grassmanian models}
A Grassmanian model in $2+1$ dimensions is defined by the equation
\begin{equation}\label{greq}
[\partial_{\mu}\partial^{\mu}P,P]=0,
\end{equation} 
where $P$ is a map from $\mathbb{R}^{2+1}$ into the 
Grassmanian manifold $Gr(m,n+1)$ of complex $m$--dimensional linear 
subspaces in $\C^{n+1}$. We shall think of $P$ as a 
complex Hermitian matrix of rank $m$ such that $P^2=P$.

The field equation (\ref{greq}) can be derived from the action
\begin{equation}
\label{G_action}
S=2\int_{[t_1,t_2]\times \mathbb{R}^2}{\rm{Tr}}\left({\mathbb{P}\wedge 
\star \mathbb{P}}\right) \ , 
\end{equation}
where $\mathbb{P}=P_{\mu}dx^{\mu}$ . 
For $m=1$ the Grassmanian models  reduce to the $\mathbb{CP}^n$ models.\par
\subsection{Moduli space approximation}
All finite energy static solutions to (\ref{wardeq}) can be factorised
in terms of maps $P_{(\alpha)}$ of $\mathbb{R}^2$ into Grassmanian
manifolds 
\cite{U89,Zak}
\begin{equation}\label{uniton_decomposition}
J=K({\bf{1}}-2P_{(1)})({\bf{1}}-2P_{(2)})...({\bf{1}}-2P_{({\cal N})}) \ ,
\end{equation} 
where $K$ is a constant unitary matrix, $P_{(\alpha)}$ satisfy some
first order PDEs, and ${\cal N}\leq n$ is 
the so called uniton number. It can happen that all uniton 
factors can be shrunk
into the form
\be\label{embedding}
J=K({\bf{1}}-2P) \ ,
\ee
where $P$ maps $\mathbb{R}^2$ into some Grassmanian manifold but does not
necessarily satisfy the first order PDEs involved in the definition of
unitons. It can be easily chacked that (\ref{embedding}) is a solution
to chiral model if and only if $P$ is  a solution to Grassmanian
model. We will call such solutions {\it{Grassmanian embeddings}}. Note
that they can represent one--uniton as well as particular multi--uniton
solutions. One--uniton solutions correspond to $P$ being
(anti)instanton solution, which at the level of the Grassmanian model 
minimises the value of energy in its
topological sector. For such solutions the energy is proportional to
the topological charge of the Grassmanian projector (given by the
formula (\ref{ladtop}) in Section \ref{section3}). This is also true
for the potential energy of the chiral field $J$, 
defined in  (\ref{energy}), since
in the case of (\ref{embedding}) it is equal to the energy of $P$. 

Integrability of the model enables a construction
of time dependent solutions by twistor and inverse scattering methods
\cite{W88,W90}. Approximate solutions corresponding to low energy exact
solutions can also be sought by a modification of Manton's 
geodesic approximation. The modification relies on taking into account a
background magnetic field in the moduli space of static solutions
induced by WZW term in (\ref{S}), and has been discussed in \cite{DM05} for
the $SU(2)$ models. In this reference the moduli space has been 
constructed from static solutions of the
model obtained by embedding 
the instanton solutions of the $\mathbb{CP}^1$
model  (which together with an analogous procedure of embedding anti-instantons
 gives all static solutions in the  $SU(2)$ case). 
It has been demonstrated that the
magnetic field vanishes and so the integrable $SU(2)$ chiral model
appears 
to be equivalent to the usual $SU(2)$ chiral model  at the level of
the 
approximation. The proof given in \cite{DM05} relied on the fact
that $SU(2)$ is three dimensional, and 
it remained uncertain
how the magnetic field  behaves for higher dimensions of the target
manifold. The main result of this paper is to clarify this and to
show that the magnetic field 
vanishes on moduli spaces constructed from the Grassmanian embeddings 
into $U(n+1)$ models for arbitrary $n$. The static Grassmanian
solutions  will not be required to be 
instantons or anti-instantons in our proof. 
\section{The metric and the magnetic field on the moduli space}
\label{section2}
\setcounter{equation}{0}
The boundary conditions (\ref{fecond}) imply that the
finite energy static solutions to (\ref{wardeq}) are maps from $S^2$ 
(conformal compactification of $\mathbb{R}^2$) into $U(n+1)$. 
In the moduli space approximation we choose a class 
of such solutions which are homotopic as maps of $S^2$ 
into $U(n+1)$ and all have the same value of potential energy.
Ideally every such map ought to provide minimum
of  the potential energy. This is the case on the level of the Grassmanian
models for constructions which involve (anti)instanton solutions. For
chiral models one can show that
all finite energy static solutions are saddle
points of the potential
energy functional \cite{PSZ88}. This raises a question about stability
of the approximate solutions. 

For a given value of the topological
charge,  all solutions in the class can be
described by finite set of parameters, which in the case of instantons 
are positions of zeroes and poles of holomorphic functions.
To ensure finite values of kinetic energy we
need to 
impose the {\it{base condition}} on the solutions by fixing their
value at
 spatial infinity. Then the parameters, if chosen appropriately, may
 define 
a map on the resulting moduli space. 
Next we allow the parameters to depend on time and so time dependent 
approximate solutions correspond to paths in the moduli space. Let us 
denote the solutions contributing to the moduli space by
$J(\gamma;x,y)$, 
where $\gamma$ denote real parameters. Approximate time dependent 
solutions are then of the form $J(\gamma(t);x,y)$ and time 
differentiation gives
\begin{equation}\label{trule}
J_t=J_j\dot \gamma^j \ , \qquad j=1, ..., \mbox{dim}\;{\cal M}.
\end{equation}
The dynamics is governed by the action obtained as a restriction of
(\ref{S})  to the moduli space
\begin{equation}
\label{effectiveS}
S_{\cal M}=\int_{t_1}^{t_2}\left({\frac{1}{2}h_{jk}\dot \gamma^j \dot \gamma^k
    +A_j  \dot \gamma^j}\right){\rm{\>dt}} \ .
\end{equation}
The metric term can be obtained from kinetic energy form
(\ref{energy}) by  use of (\ref{trule})
\begin{equation}\label{metric}
T=\frac{1}{2}h_{jk} \dot \gamma^j \dot \gamma^k \ , \ \ \ \ \ \
h_{jk}= -\int {\rm{Tr}}\left({J^{-1}J_j J^{-1}J_k}\right){\rm{\>dx\>dy}} \ ,
\end{equation}
and the magnetic term can similarly be obtained from the WZW-term, which
can be  rewritten by cyclic property of the trace as
\begin{eqnarray*}
S_M&=&\int_{t_1}^{t_2}\int_{\mathbb{R}^2} \int_0^1 {\rm{Tr}}([\hat
J^{-1}\hat  J_t, \hat J^{-1}\hat J_y] \hat J^{-1}\hat
J_{\rho}){\rm{\>d\rho \>dx\>dy}}{\rm{\>dt}}
\nonumber \\ 
&=&\int_{t_1}^{t_2}A_j \dot \gamma^j{\rm{\>dt}}, 
\end{eqnarray*}
where
\be
\label{Ajgj}
A_j=
\int_{\mathbb{R}^2} \int_0^1 {\rm{Tr}} ([\hat J^{-1}\hat J_j, 
\hat J^{-1}\hat J_y] \hat J^{-1}\hat J_{\rho}){\rm{\>d\rho\>dx\>dy}} \ .
\ee
Then $A=A_j d\gamma^j$ is the {\it magnetic 1-form} on the moduli space. 
We shall now prove the following
\begin{theorem}\label{tznikaniepola}
The magnetic field (\ref{Fij}) vanishes on moduli spaces constructed 
from embeddings (\ref{embedding}) of Grassmanian solutions. 
\end{theorem}
\begin{proof}
The essence of WZW term is that its variation does not depend on the 
particular choice of the extension $\hat J$. We consider the
variations restricted to the moduli space
$\delta J=J_i \delta \gamma^i$, and find
\begin{eqnarray*}
\delta S_M&=&\int_{t_1}^{t_2}\int_{\mathbb{R}^2}{\rm{Tr}}
\left({J^{-1} J_y \> [J^{-1}\delta J,J^{-1}J_t]}\right) \
{\rm{dx\>dy\>dt}}  \nonumber,\\
&=&-\int_{t_1}^{t_2}\int_{\mathbb{R}^2}{\rm{Tr}}\left({J^{-1}J_y \>
    [J^{-1}J_i,J^{-1}J_j]}\right) {\rm{\>dx\>dy}} \ \dot \gamma^j
\delta  \gamma^i{\rm{ \ dt}}.
\end{eqnarray*}
Comparing this expression with the variation of
(\ref{Ajgj}) 
\[
\delta S_M=\int_{t_1}^{t_2}F_{ij}\dot \gamma^j
\delta\gamma^i{\rm{\>dt}} 
\ , \ \ \ \ F_{ij}=\partial_i A_j-\partial_j A_i \ 
\]
gives
\begin{equation}\label{Fij}
F_{ij}=-\int_{\mathbb{R}^2}{\rm{Tr}}\left({J^{-1} J_y \> 
[J^{-1}J_i,J^{-1}J_j]}\right) {\rm{\>dx\>dy}} \ ,
\end{equation}
where $F=\frac{1}{2}F_{ij}(\gamma)d\gamma^i\wedge d\gamma^j$ is the
{\it{the magnetic field}}. We can see, that
although  the 
magnetic 1-form $A$ in general depends on the choice of the extension
$\hat J$, its exterior derivative $F$ does not. Changing the
extension merely corresponds to a gauge transformation of $A$.
 
Note that the potential energy term has not been included in the
effective action (\ref{effectiveS}). The potential is proportional to
the topological charge (\ref{ladtop}), and does not contribute to the 
effective equations of motion.

Let now us consider a Grassmanian projector  $P$ depending smoothly 
on some set of variables, which we shall denote by $a,b,c$. 
From idempotency and the Leibniz rule we deduce
\[
P_a=P_a P+P P_a \ ,
\]
and 
\[
\begin{aligned}
&P_a P_b P_c=P P_a P_b P_c+P_a P P_b P_c=P P_a P_b P_c
+P_a P_b P_c-P_a P_b P P_c=\\
=&P P_a P_b P_c+P_a P_b P_c-P_a P_b P_c+P_a P_b P_c P=P P_a P_b P_c
+P_a P_b P_c P \ .
\end{aligned}
\]
Taking the trace of the above expression gives
\begin{equation}\label{form}
{\rm{Tr}}(P_a P_b P_c)={\rm{Tr}}(2 P P_a P_b P_c) \ .
\end{equation}
If $J$ is given by (\ref{embedding}) then
\begin{equation}
\label{proportional}
{\rm{Tr}}\left({J^{-1} J_y \> [J^{-1}J_i,J^{-1}J_j]}\right)\sim
{\rm{Tr}}
\left({(1-2P)P_y\> [P_i,P_j]}\right) \ ,
\end{equation}
where we have assumed  
that $K$ does not depend on parameters $\gamma$ on the moduli 
space to ensure the finiteness of the kinetic energy.
The RHS of (\ref{proportional}) vanishes because of  (\ref{form}), 
which in turn implies the  vanishing of the magnetic field (\ref{Fij}).
\end{proof}

\section{Canonical structures on the moduli space}
\label{section3}
\setcounter{equation}{0}
The $\mathbb{CP}^n$ models have been discussed in detail 
within the moduli space approach \cite{W85,SZ87}. It is convenient to choose 
a map and perform calculation in a local framework. We can represent
complex directions in $\mathbb{C}^{n+1}$, which are the elements
of  $\mathbb{CP}^n$, by vectors in $\mathbb{C}^{n+1}$ with their first
component fixed to $1$. Then the map $f$ defined by
\begin{equation}
\mathbb{CP}^n\ni (1,f^1,\dots,f^n)
\longrightarrow (f^1,\dots,f^n)\in \mathbb{C}^n
\end{equation}  
belongs to the maximal holomorphic atlas of $\mathbb{CP}^n$. The
results do not depend on the choice of this  map.
The topological charge for the $\mathbb{CP}^n$ models is 
\begin{equation}\label{ladtop}
Q=-i\int_{\mathbb{R}^2}{\rm{Tr}}(P[P_x,P_y]){\rm{\>dx\>dy}}
=-\frac{1}{4}\int_{\mathbb{R}^2}P^{*}\Phi \ ,
\end{equation}
where 
\begin{equation}\label{Fform}
\Phi=-4i \> \partial\bar\partial \> {\rm{ln}}(1+\sum_{l=1}^n
|f^{l}|^2)=-4i
\dfrac{\delta^{jk}\left({1+\sum \limits_{l=1}^n
      |f^{l}|^2}\right)-f^{j}
\bar f^{k}}{\left({1+\sum \limits_{l=1}^n
    |f^{l}|^2}\right)^2}df^{k}\wedge 
d\bar f^{j} \ 
\end{equation}
is the K\"ahler form of the Fubini-Study metric on 
$\mathbb{CP}^n$ and $P^{*}\Phi$ denotes its pull-back. 
The first expression for $Q$ given in (\ref{ladtop}) is often
more convenient for calculations, while the second clarifies 
the topological character. The matrix $P$ is given in terms of $f^l$
by
\be
\label{Vkolumna}
\begin{aligned}
P=\frac{ W\otimes W^{\dag}}{W^{\dag}W} \ , \ \ \ \ W=
\left(   
   \begin{array}{cccc}
   1     \\
  f^1   \\
   \vdots \\
   f^n
   \end{array}
\right) \ .
\end{aligned}
\ee
The equality (\ref{ladtop})
is proved by establishing that 
in the chosen map both expressions give 
\begin{equation}\label{cofniecie}
-i \int_{\mathbb{R}^2} \sum \limits_{k,j=1}^n
\dfrac{\delta^{kj}(1+\sum 
\limits_{l=1}^n |f^l|^2)-\bar f^k f^j}{(1+\sum \limits_{l=1}^n
|f^l|^2)^2} \ 
\dfrac{\partial(f^k,\bar f^j)}{\partial(x,y)}{\rm{\>dx\>dy}} \ .
\end{equation}
The most natural choice of the family of static solutions for the
purpose of construction of the moduli space is to consider solutions
which minimise the energy for a given value of topological
charge. These instanton (or anti-instanton) solutions 
correspond to $f^l, \ l=1\dots n$ being rational holomorphic 
(respectively antiholomorphic) functions of the complex variable
$z=x+iy$. 
Let us concentrate on instantons, in which case \cite{Zak}
\begin{equation}
\label{degre_Q}
Q=2\pi N \ ,
\end{equation}
where $N={\rm{max}}_{l}\>(k_{alg}f^l)$ is an integer. Here $k_{alg}f^l$ is
the algebraic degree of the rational function $f^l$. Note that
(\ref{degre_Q}) holds for any  smooth map 
$P:S^2\longrightarrow\mathbb{CP}^n$ with $N$ being the homotopy class
under the standard
isomorphism $\pi_2(\CP^n)=\Z$. To see it (e.g \cite{BT82}) consider the 
homology group  $H_2(\CP^n)$. This 
is isomorphic to $\Z$.  If $P:S^2\longrightarrow  \CP^n$ is a map from
the compactified space  to $\CP^n$, representing a homology class
$P_*[S^2]$,  we  obtain the corresponding integer by evaluating
$P_*[S^2]$  on a standard generator for $H^2(\CP^n)$ represented
by the Kahler form $\Phi$. In terms of differential forms, evaluating a
cohomology class on a homology class just means integrating, so the
evaluation of $P_*[S^2]$ on $\Phi$ is given by the RHS of 
(\ref{ladtop}).
Now consider the Hurewicz
homomorphism from $\pi_2(\CP^n)$ to $H_2(\CP^n)$  sending the homotopy 
class of  $P: S^2\longrightarrow \CP^n$ to $P_*[S^2]$, where 
$[S^2] \in H_2(S^2)$ is the fundamental class. The projective space $\CP^n$
is simply connected,  so this is  an isomorphism 
$\pi_2(\CP^n)=H_2(\CP^n) = \Z$.

For a given $N$, 
the finiteness of the energy requires the base condition to be
imposed. We therefore  fix the limit of each $f^l$ at the spatial
infinity. 
Let
us choose this limit to be equal to one for all functions $f^l$. 
Then they are of the form 
\begin{equation}\label{M_Nmap}
\begin{aligned}
f^l=\frac{p_l(z)}{q_l(z)}=\frac{(z-q^{l, 1})\dots
  (z-q^{l, N})}{(z-q^{l, N+1})
\dots (z-q^{l, 2N})} \ , \ \ \ \ l=1, \dots, n \ ,
\end{aligned}
\end{equation}
and complex numbers $q$ are holomorphic coordinates on
a finite--dimensional moduli space ${\cal M}_N\subset\cal M$. 
We can define the metric as a restriction of kinetic
energy form to ${\cal M}_N$ (like in (\ref{metric})). Its completeness is
obviously equivalent to the requirement that the kinetic energy is finite
along all curves in ${\cal M}_N$. Although the base condition was necessary to
ensure finite kinetic energies, it appears not to be sufficient as  
the metric is complete only on leafs of appropriate foliation of
${\cal M}_N$ \cite{W85,SZ87} and we need to restrict the dynamics to these
leafs. 
These restrictions are assumed to hold in
the rest of the paper and we will often use 
the symbol ${\cal M}_N$ to denote some particular leaf.\par
The metric described above, which can be obtained
explicitly from (\ref{metric}) by use of (\ref{embedding}), is
${\rm{K\ddot ahler}}$ with respect to the natural complex structure
induced by map (\ref{M_Nmap}), with the K{\"a}hler
potential
\begin{equation}
\Omega=8\int_{\mathbb{R}^2}{\rm{ln}}\sum_{l=1}^n(|p_l|^2+|q_l|^2)
{\rm{\>dx\>dy}} \ .
\end{equation}

As noted in \cite{R88} this metric can also be defined canonically. Let
$\gamma$
denote the set of real parameters of all rational functions
 (\ref{M_Nmap}), which provide real coordinates  on ${\cal M}_N$. For example
 one may consider  $\gamma=\left({(q+\ov{q})/2, (q-\ov{q})/(2i)}\right)$. In the following,
 $\gamma$ will also denote a point in ${\cal M}_N$. Let
 $\{P(\gamma; \ \cdot \ ):\mathbb{R}^2\longrightarrow \mathbb{CP}^n, \
 \gamma\in {\cal M}_N\}$ be instanton solutions of the model. 
We can define the maps
\begin{equation}
F:{\cal M}_N\times \mathbb{R}^2\longrightarrow \mathbb{CP}^n \ , \ \ \ \
F(\gamma,p)
:=P(\gamma;p) \ ,
\end{equation}
\begin{equation}\label{defFp}
\begin{aligned}
& F_p:=F( \ \cdot \ ,p):{\cal M}_N\longrightarrow \mathbb{CP}^n \\
& F_{\gamma}:=F(\gamma, \ \cdot \ ):\mathbb{R}^2\longrightarrow 
\mathbb{CP}^n \ .
\end{aligned}
\end{equation}
For each smooth vector field on ${\cal M}_N$
\begin{equation*}
X:{\cal M}_N\longrightarrow T {\cal M}_N \ , \ \ \ \ \ \ 
X\in T_{\gamma}{\cal M}_N
\end{equation*}
we can now define the metric $h$ canonically by 
\begin{equation}\label{kh}
h(X, X)=\int_{\mathbb{R}^2}\hat
h\left({F_{p*}X, F_{p*}X }\right){\rm{\>dx\>dy}} \ ,
\end{equation}
where $F_{p*}X$ denotes push-forward of a vector field, $\hat h$ is
a Fubini-Study metric on $\mathbb{CP}^n$ and integration is performed
with respect to  $p\equiv (x,y)$.

\vskip 0.1 in
Let us now observe that all the results discussed here for
$\mathbb{CP}^n$ models can be extended to  the  chiral
models. To see it consider the moduli space constructed from
$\mathbb{CP}^n$ embeddings (\ref{embedding}), 
where for convenience we set $K=-i{\bf{1}}$ :
\begin{equation}\label{Embedding}
J=-i({\bf{1}}-2P) \ .
\end{equation}
Since the kinetic energy for chiral models rewritten in terms of $P$
is precisely the one for $\mathbb{CP}^n$ models, the ${\rm{K\ddot
    ahler}}$ structure is also the same. Thus the moduli space ${\cal M}_N$ can
be considered as an arena for slowly moving $\mathbb{CP}^n$ Skyrmions,
as well as the low energy solutions to the $U(n+1)$ Ward model. 
The magnetic 1-form
(\ref{Ajgj}) is an interesting object in spite of the fact that it
does not influence the motion. In Theorem \ref{trownowaznosc} we shall
show how it arises canonically on the moduli space. Let us first make
some comments about the extensions of $J$ used in the variational
principle.

In general any $J$ can be extended, as the obstruction group 
$\pi_2(U(n+1))$ vanishes. In the case of soliton solutions to
(\ref{wardeq}) we can be more explicit. It has been shown in \cite{Dt04} that
all solitons factorise as $J=\prod_{\alpha}M_\alpha$
into a finite number of the time--dependent unitons of the
form $M_\alpha={\bf 1}-(1-e^{2i\phi_\alpha})P_\alpha$, where 
$P_\alpha=P_\alpha(x, y, t)$ are  hermitian projectors, and 
the real constants $\phi_\alpha$ are the phases of the poles on the
spectral plane. Any of these
projectors can be extended by
\be
\label{diff_ext}
M_\alpha\longrightarrow \hat{M}_\alpha=
{\bf 1}-(1-e^{2i\rho\phi_\alpha}) P_\alpha,
\ee
thus giving the extension $\hat{J}=\prod_\alpha\hat{M}_\alpha$. 
In the next Theorem we shall use an extension
\begin{equation}\label{Rozszerzenie}
\hat J(t,x,y,\rho)=\cos g(\rho){\bf{1}}+\sin g(\rho)J(t,x,y).
\end{equation}
This extension corresponds to $\mathcal{F}(J,x,y,\rho)=\cos
  g(\rho){\bf{1}}+
\sin g(\rho)J$,
however the domain of $\mathcal{F}$ should be restricted from
to \ $\mathcal{U}\times
\mathbb{R}^2 \times [0,1]$, where
\[\mathcal{U}:=\{J=-i({\bf{1}}-2P):P\in \mathbb{CP}^n\}.\] Such
restriction is allowed, since all mappings $J$ within the moduli space
take values in 
$\mathcal{U}\subset U(n+1)$.
In the
case of static solutions (\ref{Embedding}) the extensions
(\ref{diff_ext}) and (\ref{Rozszerzenie}) differ only by an overall
factor depending on $\rho$, which does not contribute to the magnetic
one form.
\begin{theorem}\label{trownowaznosc}
The magnetic 1-form (\ref{Ajgj}) induced by WZW term for the extension
(\ref{Rozszerzenie})
coincides with the canonical 1--form on ${\cal M}_N$ defined by 
\begin{equation}\label{definicjakanA}
A_{kan}(X)=\frac{\pi}{2}\int_{\mathbb{R}^2}\hat
h\left({F_{{\gamma}*}V,F_{p*}X}\right){\rm{\>dx\>dy}} \ .
\end{equation}
where $V=\p/\p x$ is the 
unit vector defining the Ward equation (\ref{wardeq}).
\end{theorem}
\begin{proof}
Let us compare components of both one forms in the map $\gamma$. From
(\ref{definicjakanA}) 
we find
\[
\begin{aligned}
(A_{kan})_j&=A_{kan}\left({\frac{\partial}{\partial \gamma^j}}\right)
=\frac{\pi}{2}\int_{\mathbb{R}^2}\hat h\left({F_{\gamma *} V(p)\ , \
    F_{p*}\> \frac{\partial}{\partial
      \gamma^j}}\right){\rm{\>dx\>dy}}\\
&=\frac{\pi}{2}\int_{\mathbb{R}^2}\Phi\left({\mathcal{J}\> F_{\gamma *} V(p)\ , \
    F_{p *}\> \frac{\partial}{\partial
      \gamma^j}}\right){\rm{\>dx\>dy}} \ ,
\end{aligned}
\]
where $\mathcal{J}$ is the standard complex structure on
$\mathbb{CP}^n$ given by
$\mathcal{J}\frac{\partial}{\partial f^l}=i\frac{\partial}{\partial
 f^l}$. 
Since
\[
\mathcal{J}\> F_{\gamma *} V(p)=i \frac{\partial f^k}{\partial
  z}\frac{\partial}{\partial f^k}-i \frac{\partial \bar f^k}{\partial
  \bar z}\frac{\partial}{\partial \bar f^k} \ , \ \ \ \ F_{p *}\>
\frac{\partial}{\partial \gamma^j}=\frac{\partial f^k}{\partial
  \gamma^j}\frac{\partial}{\partial f^k}+\frac{\partial \bar
  f^k}{\partial \gamma^j}\frac{\partial}{\partial \bar f^k} 
\ ,
\]
we can use linearity and antisymmetry of the Fubini--Study K\"ahler
form
 $\Phi$ given by (\ref{Fform}) to obtain
\[
(A_{kan})_j=\frac{\pi}{2}\int_{\mathbb{R}^2}i\left\{{f^k_z \bar f^l_j
    \ \Phi \left({\frac{\partial}{\partial
          f^k},\frac{\partial}{\partial \bar f^l}}\right)-\bar f^k_z
    f^l_j \ \Phi \left({\frac{\partial}{\partial \bar
          f^k},\frac{\partial}{\partial
          f^l}}\right)}\right\}{\rm{\>dx\>dy}} \ ,
\]
where the short notation $\partial f^k / \partial z=f^k_z,  \ \partial
f^k / \partial \gamma^j=f^k_j$ has been used. Since $\Phi$ is a
fundamental form of a hermitian metric, we have
\[
\Phi \left({\frac{\partial}{\partial \bar
      f^k},\frac{\partial}{\partial f^l}}\right)=\overline{\Phi
  \left({\frac{\partial}{\partial f^k},\frac{\partial}{\partial \bar
        f^l}}\right)} \ ,
\]
so
\[
(A_{kan})_j=-\pi \int_{\mathbb{R}^2} {\rm{Im}}\left\{{f^k_z \bar f^l_j
    \ 
\Phi \left({\frac{\partial}{\partial f^k},\frac{\partial}{\partial
    \bar f^l}}
\right)}\right\}{\rm{\>dx\>dy}} \ .
\]
Finally we use (\ref{Fform}) to obtain
\begin{equation}\label{modk}
(A_{kan})_j=\pi\int_{\mathbb{R}^2}\frac{4}{\left({1+\sum
      \limits_{r=1}^n |f^{r}|^2}\right)^2} \
{\rm{Re}}\left\{{\left({1+\sum \limits_{r=1}^n |f^{r}|^2}
\right)f^l_z \bar f^l_j-f^k_z \bar f^k f^l \bar f^l_j}\right\}
{\rm{\>dx\>dy}} \ .
\end{equation}

\vskip 0.1 in

Let us now consider the  1-form  (\ref{Ajgj}) induced by the WZW
term.  Substituting
(\ref{Rozszerzenie}) into (\ref{Ajgj}), performing $\rho$ integration
and using (\ref{Embedding}) yields
\[
A_j=-2\pi i \int_{\mathbb{R}^2} {\rm{Tr}}(P[P_j,P_y]) \ .
\]
To rewrite this expression in terms of rational functions $f^k$ one 
needs to use (\ref{Vkolumna}).
Then it is straightforward but laborious to obtain
\be
\label{pomocn}
A_j=2\pi i \int_{\mathbb{R}^2} \sum \limits_{k,l=1}^n
\dfrac{\delta^{kl}(1+\sum \limits_{r=1}^n |f^r|^2)-\bar f^k
  f^l}{(1+\sum \limits_{r=1}^n |f^r|^2)^2} \ 
\dfrac{\partial(f^k,\bar f^l)}{\partial(\gamma^j,y)}{\rm{\>dx\>dy}} \ .
\ee
For holomorphic functions $f^k$ we have
\[
\begin{aligned}\
&\dfrac{\partial(f^k,\bar f^l)}{\partial(\gamma^j,y)}=-if^k_j \bar
f^l_z-i 
\bar f^l_j f^k_z , \\
\delta^{kl}\dfrac{\partial(f^k,\bar f^l)}{\partial(\gamma^j,y)}=-2i\>
& {\rm{Re}}(\bar f^k_z f^k_j) , \ \ \ \ -\bar f^k f^l
\dfrac{\partial(f^k,\bar f^l)}
{\partial(\gamma^j,y)}=2i\>{\rm{Re}}(f^k \bar f^k_j f^l_z \bar f^l) \ .
\end{aligned}
\]
These formulae and  (\ref{pomocn}) lead to 
\begin{equation}\label{mod}
A_j=\int_{\mathbb{R}^2}\frac{4\pi}{\left({1+\sum \limits_{r=1}^n
      |f^{r}|^2}\right)^2} \ {\rm{Re}}\left\{{\left({1+\sum
        \limits_{r=1}^n |f^{r}|^2}\right)f^l_z 
\bar f^l_j-f^k_z \bar f^k f^l \bar f^l_j}\right\}{\rm{dx\,dy}}
\end{equation} 
which is the same as (\ref{modk}).
\end{proof}
Similarly, by this method we can easily prove that Ruback's metric
(\ref{kh}) is equal to the metric (\ref{metric}), obtained as a
reduction of kinetic energy form to ${\cal M}_N$.
To see it write the components of (\ref{kh})
\begin{equation}\label{hjk}
\begin{aligned}
h_{jk}&=h\left({\frac{\partial}{\partial
      \gamma^j},\frac{\partial}{\partial
      \gamma^k}}\right)=\int_{\mathbb{R}^2}\hat h\left({F_{p *}\>
    \frac{\partial}{\partial \gamma^j} \ , \ F_{p *}\>
    \frac{\partial}{\partial \gamma^k}}\right) 
{\rm{\>dx\>dy}}=\\
&=\int_{\mathbb{R}^2}\hat h\left({\frac{\partial f^r}{\partial
      \gamma^j}\frac{\partial}{\partial f^r}+\frac{\partial \bar
      f^r}{\partial \gamma^j}\frac{\partial}{\partial \bar f^r} \ , \
    \frac{\partial f^s}{\partial \gamma^k}\frac{\partial}{\partial
      f^s}+\frac{\partial \bar f^s}{\partial \gamma^k}\frac{\partial}
{\partial \bar f^s} }\right) {\rm{\>dx\>dy}}=\\
&=\int_{\mathbb{R}^2} \frac{8}{\left({1+\sum \limits_{l=1}^n
      |f^{l}|^2}\right)^2} \ {\rm{Re}}\left\{{\left({1+\sum
        \limits_{l=1}^n |f^{l}|^2}\right)f^r_j 
\bar f^r_k-f^s_j \bar f^s f^r \bar f^r_k}\right\}\> {\rm{dx\,dy}}.
\end{aligned}
\end{equation}
On the other hand the substitution 
of (\ref{Vkolumna}) into (\ref{metric}) yields
\begin{equation}
\begin{aligned}
2T&=\int_{\mathbb{R}^2} \frac{8}{(W^{\dag}W)^2} \left\{{W^{\dag}W\>
    {W^{\dag}}_t W_t-W^{\dag}W_t\> {W^{\dag}}_t W}\right\}
{\rm{\>dx\>dy}}=\\
&=\int_{\mathbb{R}^2} \frac{8}{\left({1+\sum \limits_{l=1}^n
      |f^{l}|^2}\right)^2} \left\{{\left({1+\sum \limits_{l=1}^n
        |f^{l}|^2}\right)f^r_t \bar f^r_t-f^s_t \bar f^s f^r \bar
    f^r_t}\right\}\> {\rm{\>dx\>dy}} \ .
\end{aligned}
\end{equation}

\section{Conclusions}
Chiral models in 2+1 dimensions can be made integrable by addition of
a Wess-Zumino-Witten term to the standard action. This additional
term gives rise to the magnetic 1-form on the moduli space of based
rational maps $\CP^1\longrightarrow \mathbb{CP}^n$, i.e. the space
of instanton solutions of the $\mathbb{CP}^n$ model, possibly embedded
into chiral models. The magnetic 1-form depends on the choice of extension of
the chiral field involved in the definition of the WZW term, but different 
extensions correspond to the $U(1)$ gauge transformations of the
1-form. The push forward of the space-like unit vector appearing in
(\ref{wardeq}) to the target space 
canonically defines a  1-form  on
such moduli space and we have shown that there exists a preferred gauge, 
which makes the WZW induced
magnetic 1-form equal to the one obtained 
canonically. \par
The $U(1)$ connection defined by the  magnetic 1-form is flat. This is the case
not only for moduli spaces of instanton $\mathbb{CP}^n$ solutions, as the
magnetic field vanishes on all moduli spaces constructed from
Grassmanian embeddings. These results  generalise the analysis of \cite{DM05}
which applies only to the target space $SU(2)$. A treatment
of the moduli spaces of non--commutative solitons in the integrable $U(n+1)$
chiral model was recently given in \cite{MLP06}, where even the
Abelian case $n=0$ leads to a non--trivial structure.

In the case of $U(2)$  model there are no more
possibilities, 
since here all static solutions are necessarily
Grassmanian embeddings. It remains to be seen whether it is possible
to construct the moduli spaces from static non--Grassmanian solutions of
$U(n+1)$ model for $n>1$, such that the field would not vanish. For
the Grassmanian embeddings the vanishing of the field is implied by
vanishing of its density, the integrand of (\ref{Fij}). It is possible
to construct a moduli space for the $U(3)$ model such that
this density does not vanish, however it seems to possess
symmetries which ensure vanishing of the integral. 
This has been checked only for a few  points in the moduli space, so 
the problem is open. 

The moduli space approach to the ordinary $\CP^n$ model in 2+1
dimensions does not approximate the true dynamics of the model. 
This has recently been shown by Rodnianski and Sterbenz
\cite{RS06} by a rigorous analysis of
the (non-integrable) equations of motion. Rodnianski and Sterbenz
have 
demonstrated that a class of  solutions must blow up in finite time
and a rate of this blow up is different than predicted by the
geodesic approximation.
The situation for the modified chiral model (\ref{wardeq}) 
is quite different, as
there exist exact solutions which are regular for all times 
\cite{W88,W95,Dt04}. For some of these solutions the
total (kinetic +potential) energy  is quantised at the classical level by
the elements of $\pi_3(U(n+1))$ \cite{DP06},  and thus for all $t$ 
the total energy is equal to the potential energy of some static 
solution which in turn is equal to the degree (\ref{degre_Q}) of some 
Grassmanian projector.
Solutions to (\ref{wardeq}) obtained in the moduli space approximation
presented in this paper
have energies close to their potential energy  as their
kinetic energy is small. We should therefore expect that some of these
approximate solutions arise from exact solutions by a limiting procedure.

\section*{Acknowledgements}
We  wish to thank Piotr Kosinski, Nick Manton, 
Lionel Mason  and Gabriel Paternain
for valuable discussions.
\appendix
\section{Appendix}
\setcounter{equation}{0}
The model (\ref{wardeq}) is translationally invariant and
one expects to find the conserved momentum
corresponding to the space translations. In 
\cite{W88} Ward has observed that the total energy and the
$y$--momentum  for (\ref{wardeq}) are the same as for the
ordinary chiral model, but the $x$--momentum of the chiral model is
not conserved by the time evolution (\ref{wardeq}) of the initial
data. Here we shall revisit this problem and find the $x$--momentum 
using the WZW Lagrangian
(\ref{S}) written in terms of the torsion on $U(n+1)$. 
The Lagrangian density  takes the form
\[
\mathcal{L}=-\frac{1}{2}\eta^{\mu\nu}\p_\mu \phi^i \p_\nu \phi^j g_{ij}(\phi)+\frac{1}{2}V_\alpha\varepsilon^{\alpha\mu\nu}
\lambda_{ij}(\phi)
\p_{\mu}\phi^i\p_{\nu}\phi^j \ ,
\]
where $g$ is the metric on the group, and $\lambda$ is a local
two--form potential for the totally antisymmetric torsion  \cite{W88T}.
The conserved Noether energy-momentum tensor is 
\[
T_{\mu\nu}=\eta_{\mu\nu}\mathcal{L}-\frac{\p \mathcal{L}}{\p (\p^\mu
  \phi^j)} 
\p_\nu \phi^j.
\]
The
energy corresponding to $T_{00}$ 
is given by (\ref{energy}), and the momentum densities are
\begin{eqnarray}
\mathcal{P}_y&=&T_{02}=-{\rm{Tr}}
\left({J^{-1}J_t J^{-1}J_y}\right),\nonumber\\
\label{px}
\mathcal{P}_x&=&T_{01}=-{\rm{Tr}}\left({J^{-1}J_t
J^{-1}J_x}\right)-\lambda_{ij}\p_x\phi^i \p_y\phi^j.
\end{eqnarray}
The additional term in the conserved $x$--momentum 
$P_x=\int_{\R^2}\mathcal{P}_x \d x\d y$
does not depend on the choice  of $\lambda$, since for a fixed $t$
\begin{equation}\label{dPx}
\Theta:=\int_{\mathbb{R}^2}
\lambda_{ij}(\phi)
\p_x\phi^i \p_y\phi^j
{\rm{\;dx\;dy}}=\int_{\mathbb{R}^2}J^*\lambda \ .
\end{equation}
This expression does not change under the transformation 
$\lambda\rightarrow \lambda +\d \beta$ because 
$
\int_{\mathbb{R}^2}\d(J^*\beta)=0
$
as a consequence  of the boundary condition (\ref{fecond}). 
We can therefore choose the extension  $\hat{J}$ given by (\ref{diff_ext})
to find the additional term $\Theta$ using 
the identity 
\be
\label{charge_id}
\int_{\mathbb{R}^2}\lambda_{ij}\p_x\phi^i \p_y\phi^j \;\d x\;\d
  y=\int_{\mathbb{R}^2}\int_0^1 {\rm{Tr}}\left({\hat J^{-1}\hat J_\rho
    \left[{\hat J^{-1}\hat J_y,\hat J^{-1}\hat J_x}\right]}\right)\;\d \rho\;\d
  x\;\d y,
\ee
which follows from calculating ${\cal P}_x$ in terms of $\hat{J}$ 
directly from (\ref{S}).

Consider the time--dependent 
one soliton solution \cite{W88}
\[
J=i\Big({\bf 1}-
\Big( 1- \frac{\mu}{\bar \mu} \Big) P  \Big).  
\]
Here $\mu\in\C/\R$ is a non-real constant,
$P=W\otimes W^{\dag}/||W||^2$ is the Grassmanian projection 
(\ref{Vkolumna}) and the components of $W:\R^{2, 1}\rightarrow \C^{n+1}$
are holomorphic and rational in  $\omega= x + \frac{\mu}{2} (t+y)
+\frac{\mu^{-1}}{2} (t-y) $.
In this case the additional term $\Theta$
is proportional to the topological charge (\ref{ladtop}),
which is itself a constant of motion as the time evolution
is continuous.

\end{document}